\documentclass[fleqn,twoside]{article}
\usepackage{espcrc2}

\usepackage{graphicx}

\newcommand{\AmS}{{\protect\the\textfont2
  A\kern-.1667em\lower.5ex\hbox{M}\kern-.125emS}}

\title{Observation of a Sharp Lambda Peak in the Third Harmonic Voltage Response of High-T$_c$ Superconductor Thin Films.}

\author{Nicolas Ch\'{e}enne\address[VSM]{Laboratorium voor Vaste-Stoffysica en Magnetisme,
        Katholieke Universiteit Leuven,\\
        Celestijnenlaan 200 D, B-3001 Leuven, Belgium}%
        \thanks{Grants: This research has been supported by the Belgian DWTC, IUAP, the Flemish GOA and VIS/97/01
Programmes. T.M. is KUL Senior Fellow (F/00/038) and N.Ch. is
supported by the EC TMR Network Contract nr. ERB-FMRX-CT-980171
and by the Flemish FWO project G.0222.02.},
        Todor Mishonov\addressmark[VSM]\address[Sofia]{Department of Theoretical Physics,
        Sofia University St. Kliment Ohridski,
        5 J. Bourchier Blvd., 1164 Sofia, Bulgaria}
        and
        Joseph O. Indekeu\addressmark[VSM]\thanks{\mbox{Corresponding author:~~~~~~~~}
        \mbox{phone: (+32 16) 327 127,}
        \mbox{fax.: (+32 16) 327 983,}
        \mbox{e-mail:joseph.indekeu@fys.kuleuven.ac.be}.}}

\date{\today}

\begin{document}

\begin{abstract}
In this paper, we report on the sharp peak observed in the third
harmonic voltage response generated by a bias sinusoidal current
applied to several strips patterned in a
YBa$_2$Cu$_3$O$_{7-\delta}$ thin film and in two
La$_{1-x}$Sr$_x$CuO$_4$ thin films, when the temperature is close
to the normal-superconductor transition. The lambda-shaped
temperature dependence of the third harmonic signal on the current
characteristics is studied. Several physical mechanisms of third
harmonic generation are examined. PACS Code/Keywords: 74.25.Fy -
Third harmonic response, Non-linear voltage response, high-T$_c$
Superconductor.\vspace{1pc}
\end{abstract}


\maketitle

\section{Introduction}

Harmonic signals created by ac excitation have been used for a
long time in the investigation of high temperature superconductors
(HTSC) \cite{Ji}. Particularly, a third harmonic signal is a
consequence of non-linear effects and its measurement is a
powerful technique to investigate, and moreover, to discriminate
between the different possible origins of these non-linearities.
Different techniques of third harmonic measurement exist for
investigating properties of HTSC devices especially in microwave
range \cite{Trunin}-\cite{Shimakage}, or for characterizing
material properties such as thermal conductivity
\cite{Robbes}-\cite{Lu}, specific heat \cite{Lu}, current response
\cite{Dahm}, or critical temperature \cite{Raphael98}.\\
\indent The existence of a sharp maximum in the third harmonic can
be easily understood from a qualitative point of view. At low
temperatures the voltage response of the superconducting phase is
almost zero. On the other hand, in the normal phase we have with
high accuracy an ohmic behaviour and the response at higher
frequency harmonics is again small. A sharp maximum related to a
viable variety of nonlinear phenomena is expected only close to
the critical temperature.\\
\indent As an example, measurement of the third harmonic of the ac
susceptibility revealed the role of interplanar spacing on
formation of flux lines and demonstrated that the irreversibility
field in Bi2212 is determined by surface barriers
\cite{Raphael98}, \cite{Raphael00}. This is related to vortex
motion, which is one of the possible origins of the non-linear
effects in electrical properties in HTSC \cite{Mallozi}. See also
references \cite{Hlubina} and \cite{Sandeman} where a mechanism is
discussed for the generation of third harmonics based on the anisotropic
scattering rate of normal charge carriers included in the Boltzmann equation.\\
\indent One of the other possible roots of this non-linear
behaviour is given in ref. \cite{MPI}, where a theoretical
derivation is presented to explain non-linear effects in
electrical conductivity close to the transition temperature. From
this paper \cite{MPI}, it is expected that the measurement of the
third harmonic voltage allows getting hold of superconductor
constants of the BCS theory. In addition, the authors \cite{MPI},
\cite{MCRI}, investigate theoretically the third harmonic voltage
generation produced by the thermal response of the film to a
harmonic current. Third harmonic response is definitely related to
different roots of non-linear effects and should allow one to
gather new information about the
mechanism of HTSC non-linearity. \\
\indent In the present paper, systematic measurements of the
voltage response to a purely harmonic current crossing HTSC strips
patterned in a YBa$_2$Cu$_3$O$_{7-\delta}$ (YBCO) thin film and in
two La$_{2-x}$Sr$_x$CuO$_4$ (LSCO) thin films with x=0.15 were
performed as a function of temperature. First and third harmonics
were recorded for the whole temperature range, from room
temperature to a few Kelvin below the normal-superconductor
transition. A sharp peak was observed in the third harmonic
voltage close to the critical temperature, exhibiting a
lambda-shaped ($\lambda$-shaped) behaviour. This $\lambda$-shaped
peak was investigated for different current amplitudes and frequencies. \\
\indent Sample and measurement descriptions are given in part 2 of
this paper. Complete experimental data processing and observations
are presented in section 3. In section 4, mechanisms of third
harmonic signal generation are briefly discussed.

\section{Experimental setup}
\begin{figure}[t]
\includegraphics*[height=10pc]{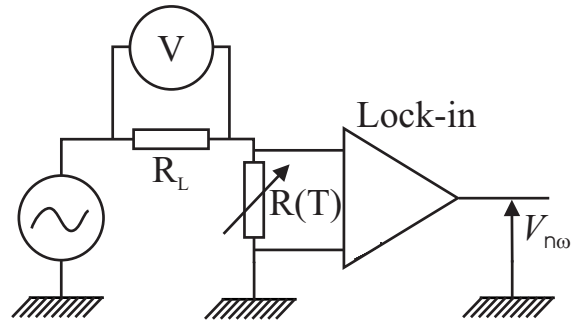}
\caption{Scheme of the electrical measurements. The current in the
superconductor sample (symbolized by its temperature-dependent
resistance $R(T)$) is governed by the voltage generator and the
load resistance $R_L$. The voltmeter across $R_L$ allows to
monitor the current in the sample. The rms voltage harmonic
response ($V_{n\omega}\div\surd2$, n~=~1,~3) across the sample is
measured by a lock-in amplifier and recorded as a function of
temperature.} \label{fig:scheme}
\end{figure}

Measurements were first performed on two different strips designed
in a YBCO thin film, 180 nm thick, deposited by DC sputtering on a
MgO substrate, as described in \cite{Wagner} and \cite{Wuyts}. The
smallest is 1.59~mm long and 77~$\mu m$ wide, and the largest is
2.87~mm long and 474~$\mu m$ wide. Both were chemically etched in
acid solution. A second set of measurements was performed on 1 mm
long and 100~$\mu m$ wide strips patterned on two different LSCO
films, LSCO1 and LSCO2, both deposited by sputtering on a
SrTiO$_3$ substrate. The first one was synthesized at
850$^\circ$C, under 1.83~mbar of pure oxygen, and annealed under
pure oxygen at 10~mbar at 850$^\circ$C for half an hour. For the
second one, the temperature of deposition and annealing was
decreased to 840$^\circ$C.\\
\indent The samples were placed in a cryostat and exposed to a
helium flux. The connection between the electronic equipment and
the strip was made via coaxial cables in a four-points geometry
(fig.~\ref{fig:scheme}). Strips were bias current, the latter
being sinusoidally modulated with amplitude $I_0$ and frequency
$f$~$=$~$\omega$$/$$2\pi$. The rms current value $I_0/\sqrt{2}$
was monitored during the whole experiment by measuring the voltage
across the load resistance $R_L$, chosen large enough to have
negligible variation of the current (less than 10~\%). Using a
digital lock-in amplifier (model SR830 from Stanford Research
Systems), first ($V_{1\omega}$) and third ($V_{3\omega}$)
harmonics of the voltage developed in the sample were successively
recorded simultaneously with temperature and time.\\
As was mentioned in previous papers \cite{Robbes}, \cite{MCRI},
the current has to be much smaller than a maximal value $I_G$:
\begin{equation}
\label{maxcurrent} I_G~=\sqrt{\frac{G}{\left|R'\right|}}~,
\end{equation}
where $R'$ is the first derivative of the electrical resistance
with respect to temperature and $G$ is the thermal boundary
conductance between the strip and the substrate, which is
maintained at the temperature of the helium bath. $G$ can be
expressed through the thermal boundary conductivity $g$, the
length $L$ and the width $w$ of the strip:
\begin{equation}
\label{G}
G=gLw
\end{equation}
In the case of the YBCO film, calculating $I_G$ from standard
values of $g$ (2000~W/Kcm$^2$) and $R'$ (1~$\Omega$/K) in the
normal state, we found the maximal current to equal 1.6~A for the
narrowest strip and 5.2~A for the widest one. In our case, the
maximum current used had an amplitude of 7~mA, far enough from the
limit given by Eq.~(\ref{maxcurrent}). Concerning the LSCO films,
the biggest current amplitude used was 1.4~mA.

\section{Experimental observations}
\subsection{Experimental Data processing}
\begin{figure}[b]
\includegraphics*[bb=0 0 600 820, height=38pc]{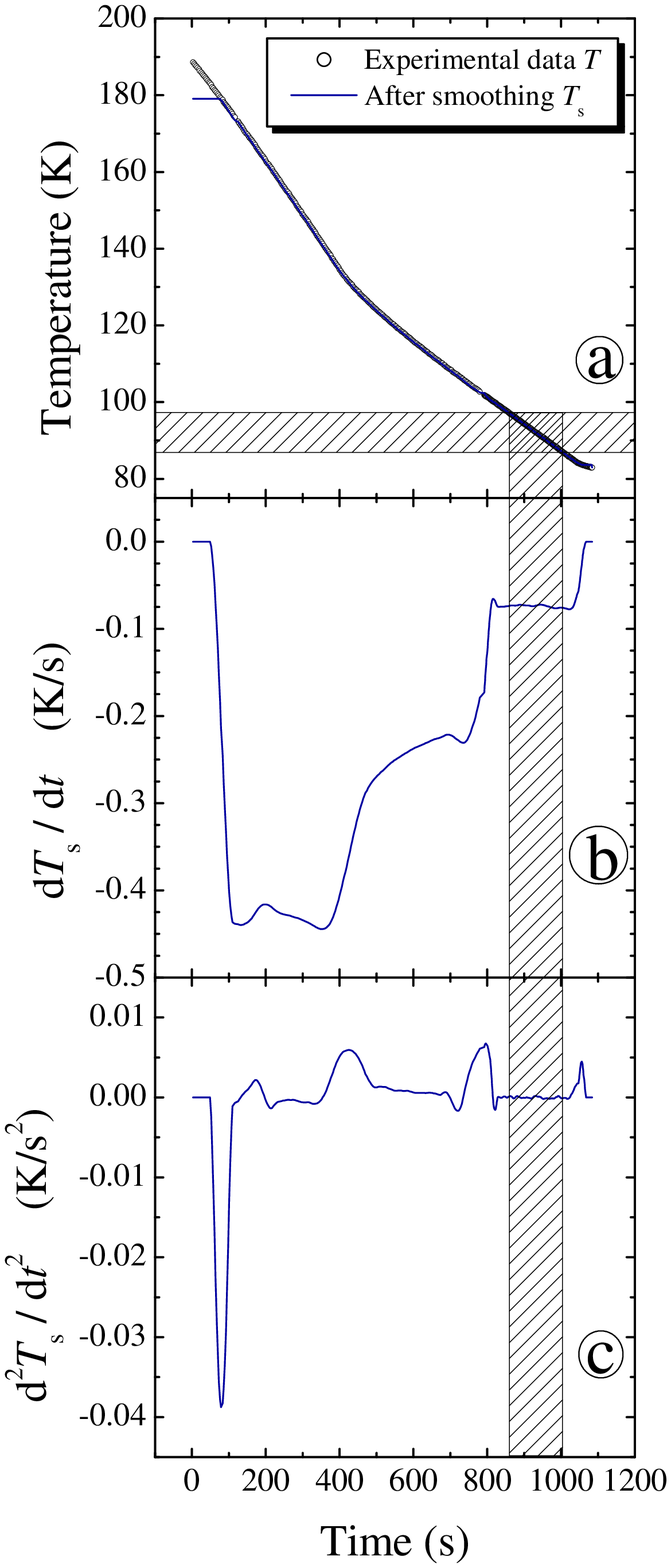}
\caption{Recording of temperature $T$ vs time (a), and first (b)
and second (c) derivatives of its smoothed values $T_s$.}
\label{fig:TempTime}
\end{figure}

\begin{figure}[b]
\includegraphics*[bb=0 0 600 820, height=38pc]{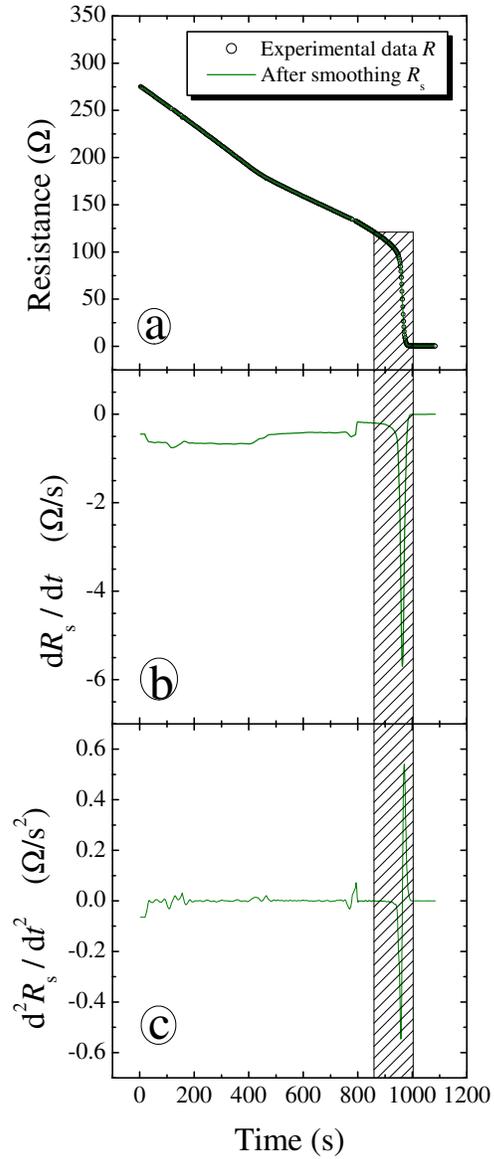}
\caption{Recording of resistance $R$ vs time (a), and first (b)
and second (c) derivatives of its smoothed values $R_s$.}
\label{fig:ResTime}
\end{figure}

\begin{figure}[b]
\includegraphics*[bb=0 0 600 820, height=38pc]{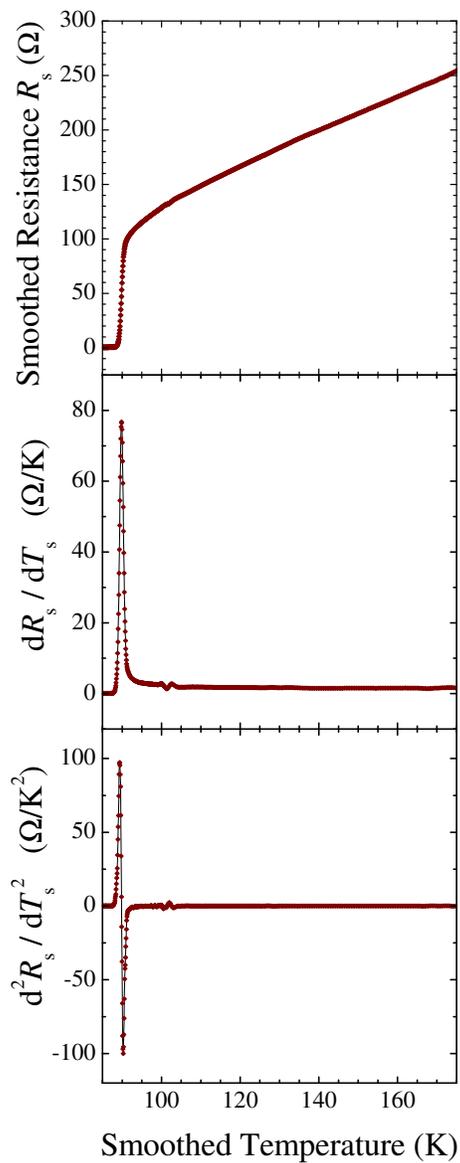}
\caption{The resistance and its first and second derivatives with
respect to temperature deduced from eq. \protect\ref{derivatives}
and data sets extracted from figures \protect\ref{fig:TempTime}
and \protect\ref{fig:ResTime}.} \label{fig:ResTempLarge}
\end{figure}

\begin{figure}[b]
\includegraphics*[bb=0 0 600 820, height=38pc]{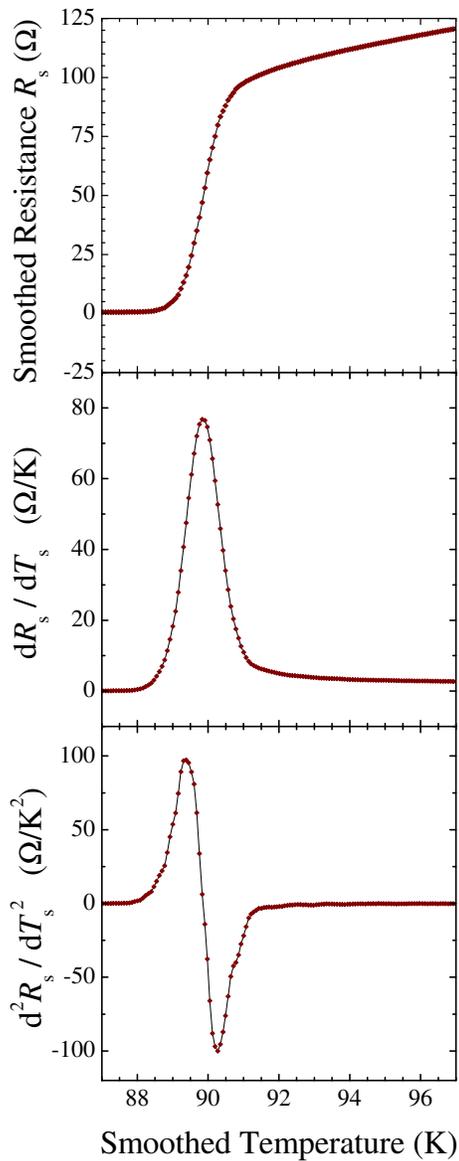}
\caption{The resistance and its first and second derivative with
respect to temperature, on the temperature range indicated by the
hatched areas on figures \protect\ref{fig:TempTime} and
\protect\ref{fig:ResTime}.} \label{fig:ResTempSmall}
\end{figure}

Before showing our results, we explain in some detail the
experimental data processing, taking as an example the
measurements obtained from the YBCO film. The goal of this
processing is to calculate the first and the second derivatives of
the resistance with respect to temperature, which we need to
estimate the thermal conductance $G$, as described in ref.
\cite{MCRI}. Since our cryogenic system induces strong temperature
noise, temperature data were smoothed over time, using a
Savitzky-Golay filter \cite{Press}, which performs a local
polynomial regression to determine the smoothed value for each
data point. In our case, $2^{nd}$ order polynomial regression was
used, taking into account 25 points to the left and 25 points to the right.\\
\indent The same smoothing was used to filter the electrical
resistance of the strip, $R$. Since the frequency $\omega$ is in
the kHz range, we consider the superconducting strip as a purely
ohmic device where inductive and capacitive components are
negligible. The resistance is then deduced from $V_{1\omega}$ by
Ohm's law:
\begin{equation}
\label{Ohm}
R_{1\omega}=\frac{V_{1\omega}}{I_0}
\end{equation}
\noindent In this case, only 5 points to the left and 5 points to
the right were taken into account to perform the local regression,
to avoid any big distortion close to the superconducting
transition. Exactly the same procedure was applied for smoothing
the third harmonic voltage.\\
\indent Thus, sets of time-dependent smoothed-data were obtained
and used to calculate the first and second derivatives of the
temperature ($\dot{T}$, $\ddot{T}$) and resistance ($\dot{R}$,
$\ddot{R}$) with respect to time $t$, differentiating using
Savitzky-Golay smoothing. For this step, the local smoothing was
performed on the smoothed-data sets. The local regression taking
into account 21 points for the temperature set and 11 points for
the resistance set, allowed to deduce the first and the second
derivatives for each data point. Thus, it was straightforward to
determine for each point in time, the values of the first and
second derivatives of the resistance with respect to temperature
$T$:
\begin{equation}
\label{derivatives} R'(T)=\frac{\dot{R}}{\dot{T}},\qquad
R''(T)=\frac{\ddot{R}\dot{T}-\dot{R}\ddot{T}}{(\dot{T})^3},
\end{equation}
\indent All the steps of the experimental data processing are
gathered in figures~\ref{fig:TempTime} and \ref{fig:ResTime},
presenting, respectively, the temperature, the resistance and
their respective derivatives obtained for the narrowest strip
excited by a current of 4~mA~rms. The temperature was shifted by a
few Kelvin during the measurements due to the poor thermal
conductivity of the sample holder on which our temperature diode
and sample were glued; we then shifted all temperature data sets
by 4~K in order to have a reasonable $T_c$ of 90~K for our YBCO sample.\\
\indent Figures~\ref{fig:ResTempLarge} and \ref{fig:ResTempSmall}
give the resistance and its first and second derivatives with
respect to smoothed temperature, deduced from the time-dependent
data sets shown in figures~\ref{fig:TempTime} and
\ref{fig:ResTime}, with use of eq.~\ref{derivatives}. These
results were used to calculate the thermal conductance $G$ between
the strip and the heat sink, as described in \cite{MCRI} and to
evaluate the contribution of the thermal effect in the peak
observed in the third harmonic voltage $V_{3\omega}$, which will
be described in the next section.

\subsection{Lambda shape of the 3$^{rd}$ harmonic voltage $V_{3\omega}$}
\begin{figure*}[thb]
\begin{center}
\includegraphics*[bb= 0 0 590 590, height=40pc]{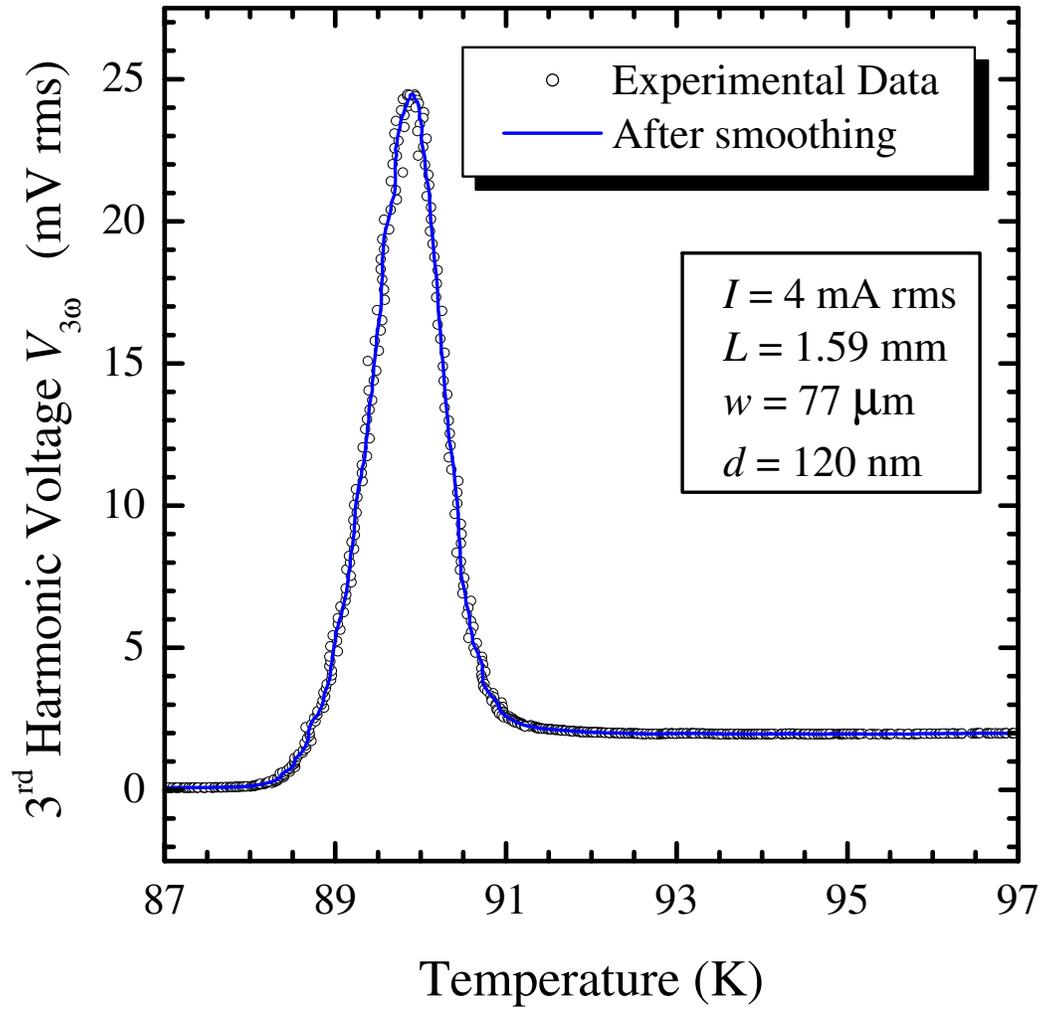}
\caption{$\lambda$-shaped temperature dependence of the third
harmonic voltage $V_{3\omega}$ for YBCO.} \label{fig:LambdaPeak}
\end{center}
\end{figure*}

\begin{figure}[b]
\includegraphics*[bb= 0 0 500 700, height=33pc]{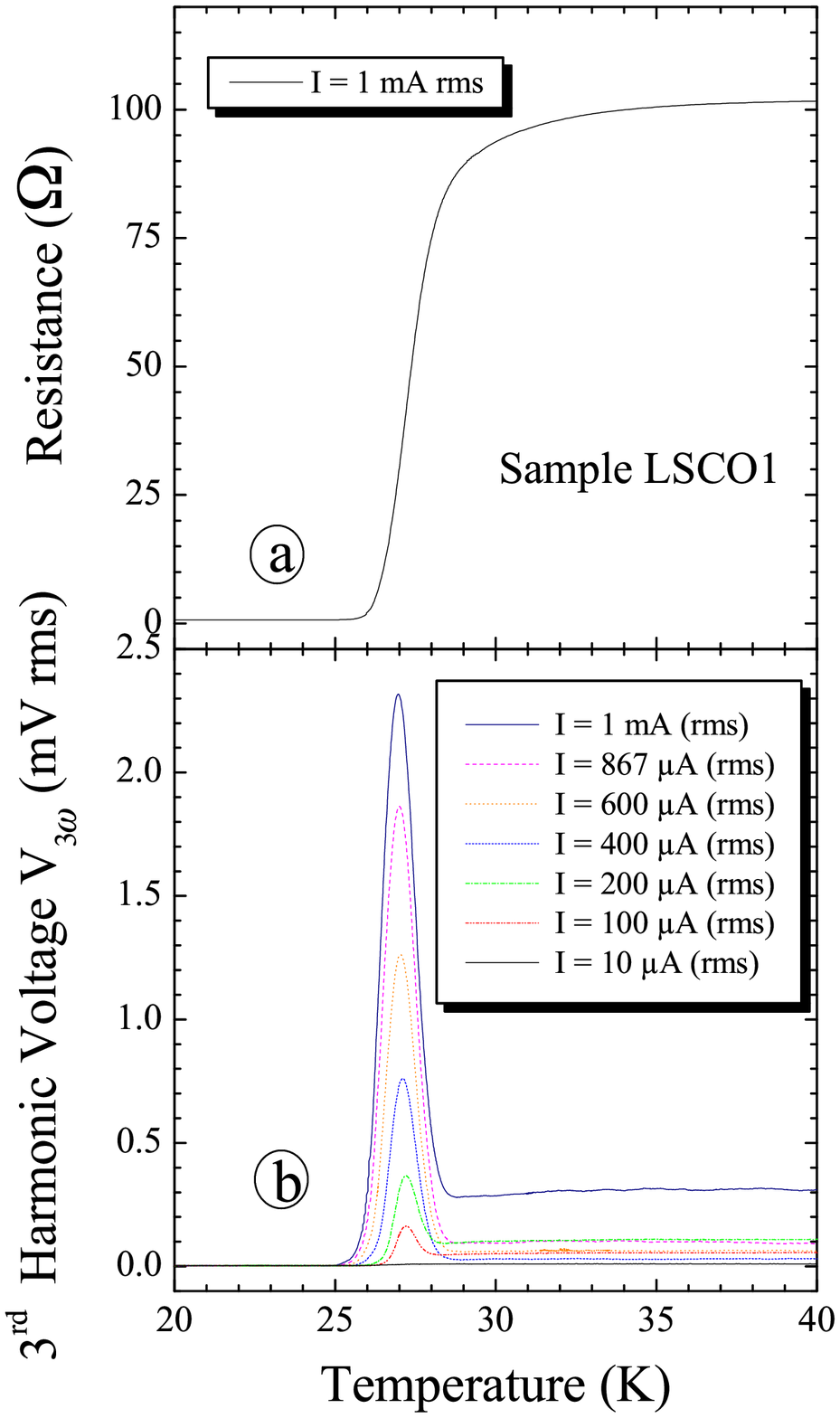}
\caption{Measurements on LSCO1.  a:~Dependence of the strip
resistance on temperature ; b:~Dependence of the third harmonic
signal on temperature for different current amplitudes.}
\label{fig:LS239_lambda_peak}
\end{figure}

\begin{figure}[b]
\includegraphics*[bb= 0 0 500 700, height=33pc]{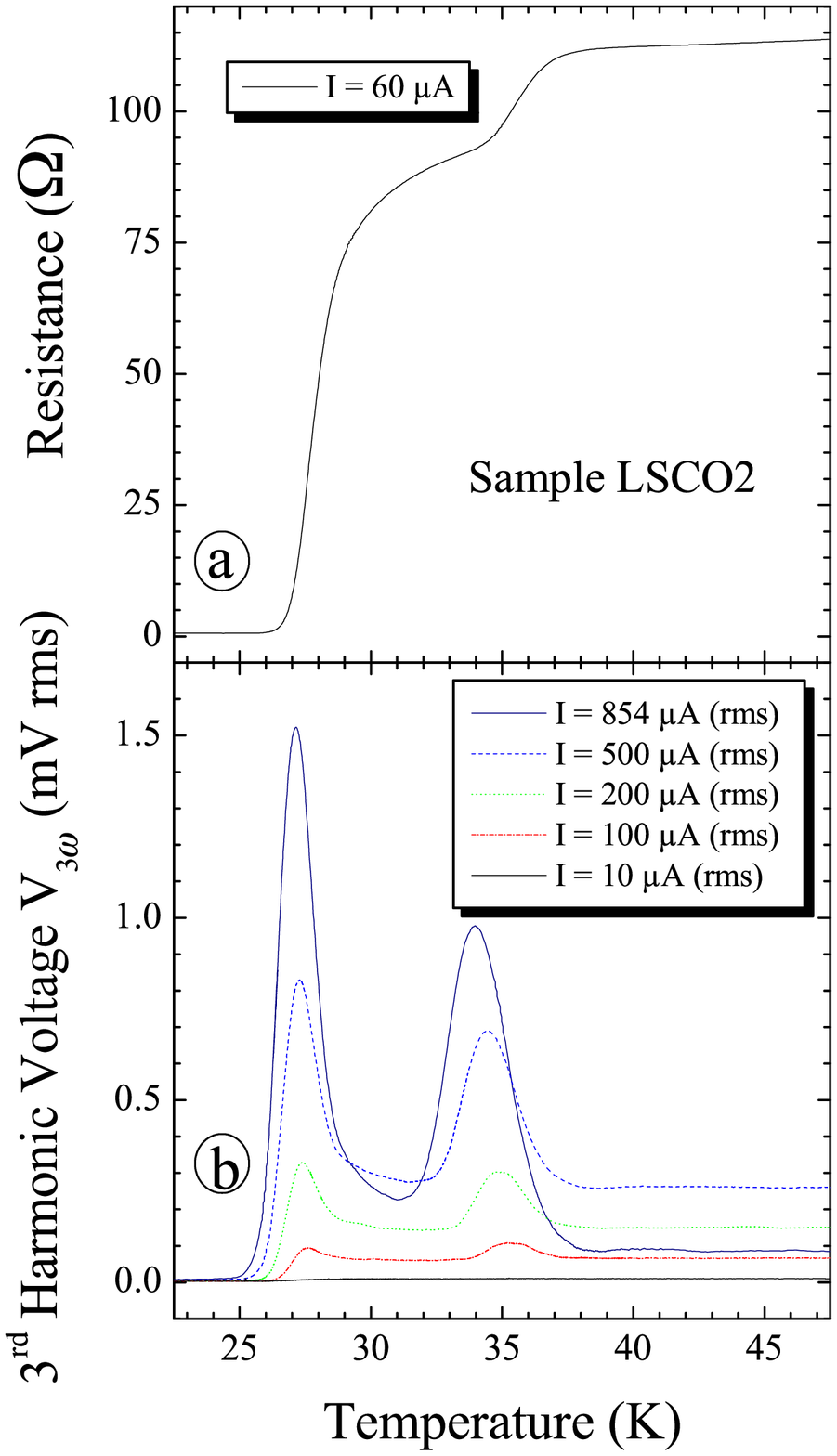}
\caption{Measurements on LSCO2.  a:~Dependence of the strip
resistance on temperature ; b:~Dependence of the third harmonic
signal on temperature for different current amplitudes.}
\label{fig:LS240_lambda_peak}
\end{figure}

\begin{figure*}[t]
\begin{center}
\includegraphics*[bb= 0 0 400 400, height=35pc]{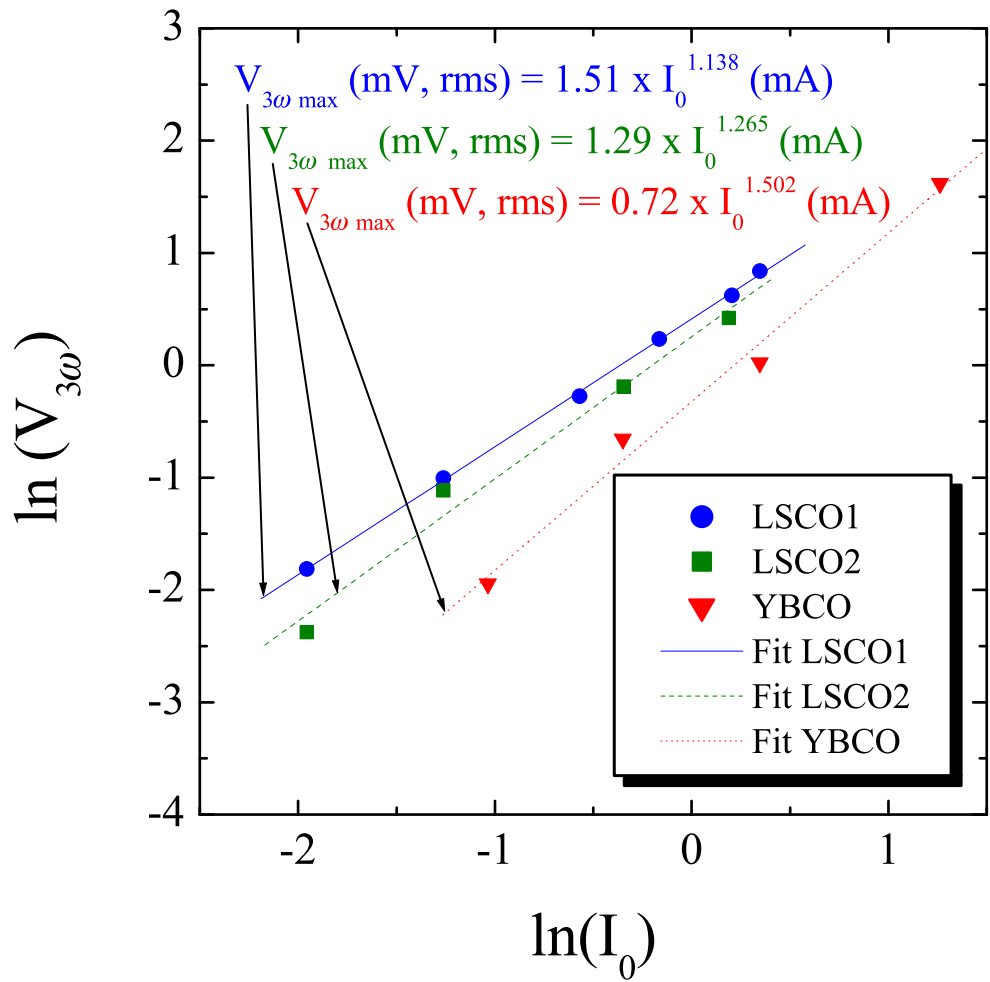}
\caption{Dependence of the maximum of the third harmonic signal on
current amplitudes for the three tested samples.}
\label{fig:Current}
\end{center}
\end{figure*}

Figure \ref{fig:LambdaPeak} gives the temperature dependence of
the third harmonic response of the narrowest strip patterned on
the YBCO film to the current excitation with amplitude $I_0$ equal
to 5.65~mA and a modulation frequency of 1024~Hz. \\
\indent To have an appreciation of the common behaviour of the
third harmonic response for different samples, we performed the
same kind of measurement on two LSCO films. Figures
\ref{fig:LS239_lambda_peak} and \ref{fig:LS240_lambda_peak}
present for each tested sample (LSCO1 and LSCO2) both the
resistance and the third harmonic response obtained for currents
ranging from 10~$\mu$ A~rms to about 1~mA~rms, as a function of
the temperature. The sharp transition of the resistance presented
in figure \ref{fig:LS239_lambda_peak}a gives rise to a unique
lambda-shaped peak in figure \ref{fig:LS239_lambda_peak}b. In the
case of a sample exhibiting a ``double" transition in the
resistivity as for LSCO2 (see fig.~\ref{fig:LS240_lambda_peak}a),
we observe two peaks in the third harmonic voltage response,
fig.~\ref{fig:LS240_lambda_peak}b. This puts into evidence the
high accuracy of this technique of measurement, firstly, to
determine the critical temperature of a material in a very simple
way, and further, to differentiate between different phases in a
material and to give for each of them the critical temperature. \\
\indent The behaviour of the shape and the amplitude of the peak
with respect to current characteristics has been explored. As we
can see on figures \ref{fig:LS239_lambda_peak}b and
\ref{fig:LS240_lambda_peak}b the amplitude of the peak increase
with the current. Figure \ref{fig:Current} presents the dependence
of the peak value of the third harmonic response on the current
amplitude in a log-log plot. A power law dependence of the third
harmonic peak amplitude on the current intensity is exhibited for
each curve. But no clear power law emerges, the power ranging from
1.1 for the LSCO1 film to 3/2 for the YBCO film. We are also not
aware of a theoretical prediction of a power law for the peak
amplitude versus current. At $T_c$, various contributions to the third
harmonic play a role and are difficult to estimate. \\
\indent For $T$~$>$~$T_c$, the situation is easier to analyse
theoretically \cite{MPI}, \cite{MCRI} and a power law with an
exponent 3 is predicted for the dependence of the third harmonic
amplitude on the current. In our experiments, however, a
complication arises due to the presence of a parasite current
coming from our generator, $I_{3\omega}$, which produces a third
harmonic signal that plays an important role in this temperature
range. This effect hides the real amplitude of the third harmonic
signal. Following the theoretical works \cite{MPI}, \cite{MCRI},
we have adopted a fitting procedure which takes into account this
parasite current. In any case, even with these difficulties, it is
important to note that the third harmonic signal presented in this
paper follows qualitatively the expected lambda-shaped behaviour
predicted in \cite{MCRI}.

\subsection{Experimental Data Analysis}
\indent Figure \ref{fig:Fit} presents the first attempt to connect
theory and experiment. It shows the fit of the lambda shape, in
the case of the 1~mA~rms current excitation, using two different
functions. The first one takes into account only the thermal
effect. In the second one, both the thermal effect and the
nonlinear current generated by the generator, $I_{3\omega}$, are
considered. \\
\indent To have an appreciation of the reliability of the fit, the
thermal resistivity $r_h$ between the film and the heat sink has
been calculated from the thermal resistance $R_h$ given by the
fit, considering the surface of the patterned strip (100~$\mu$m
$\times$ 1~mm). If we consider the case where both the parasitic
current $I_{3\omega}$ and the thermal conductance effects are
pesent, the fit gives a value of $R_h$ equal to 580~K/W, which
corresponds to a thermal resistivity equal to
5.8~$\times$~10$^{-5}$~m$^2$K/W at the considered temperature
(about 30~K). To our knowledge, no data has been published
concerning the thermal boundary resistivity of LSCO thin films at
this temperature. Then we compare the obtained value to the one
theoretically predicted in \cite{Phelan} for a YBCO thin film
deposited on MgO. The value predicted in the latter \cite{Phelan}
is about 10$^{-6}$~m$^2$K/W at 30~K. Our experimental value is not
far from this theoretically predicted value (about one order of
magnitude bigger). Because the nature of substrate and film are
different, some difference is to be expected. Furthermore, in the
range of temperatures considered (20~K~$\leq$~T~$\leq$~60~K), the
variation of the thermal boundary resistivity is rather large.
\indent The dependence of the peak on the current frequency was
studied at fixed amplitude current ($I_0$~=~1.41~mA). No
differences in the peak were observed for frequencies ranging from
114~Hz to
20.6~kHz.\\

\section{Discussion and conclusion}
\begin{figure*}[b]
\begin{center}
\includegraphics*[bb= 0 0 480 480, height=35pc]{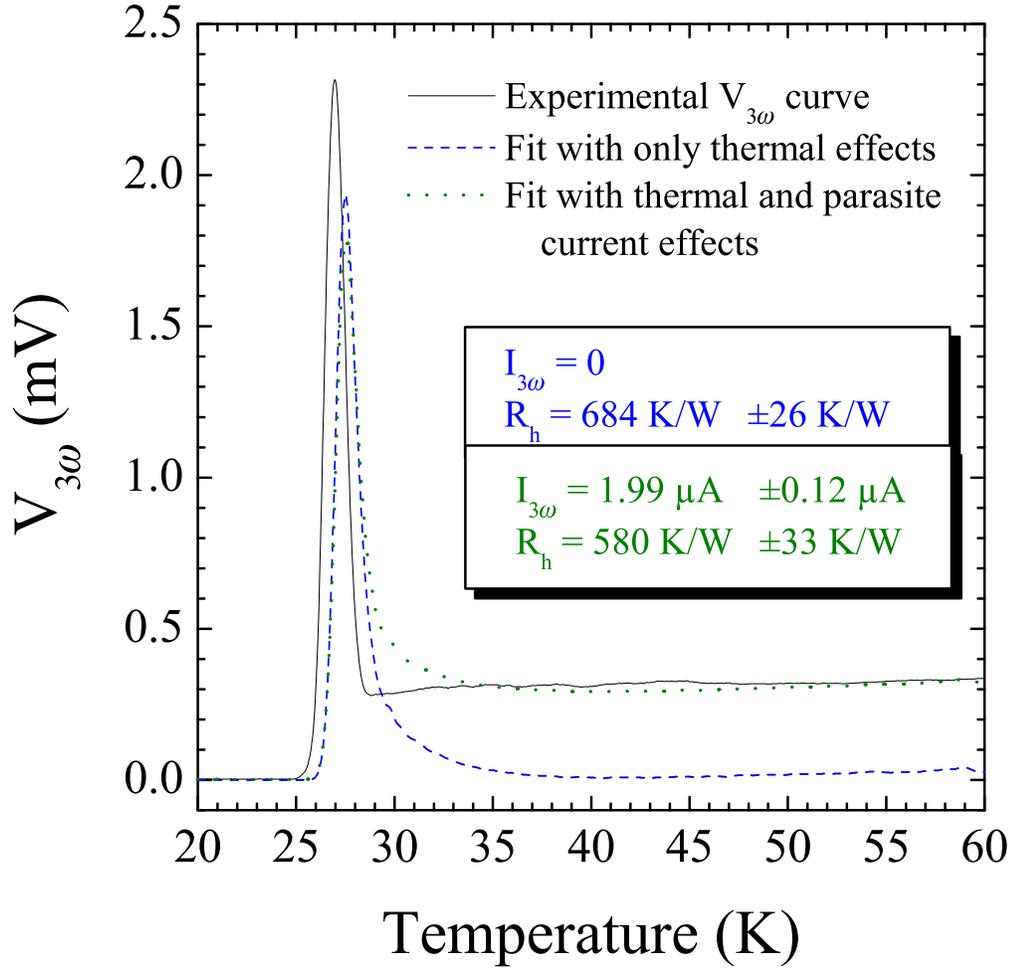}
\caption{Fit of the lambda-shaped peak in the case of a 1~mA~rms
current measurement on LSCO1, taking into account the thermal
effect only (with the thermal resistance $R_h$ between film and
substrate as parameter), or both the thermal and the parasite
$I_{3\omega}$ current effects. The equation used for the fitting
procedure is:
$V_{3\omega}=RI_{3\omega}+\frac{R_hRR'I_0^3}{4}+\frac{5R_h^2R(2R'^2+RR'')I_0^5}{32}$}
\label{fig:Fit}
\end{center}
\end{figure*}

\indent The main qualitative property of the third harmonic
response is the sharp $\lambda$-peak close to the critical
temperature. Taking into account the third harmonic signal
generated by the thermal oscillations in the film and by the
parasite current from the generator, the experimental data have
been fitted. From this procedure, we found an acceptable value of
the thermal resistivity, $r_h$, and a reasonable value of the
parasite current, $I_{3\omega}$. Thus, we have for the moment a
qualitative agreement between experiments and theory. \\
\indent A more detailed investigation, beyond the partial analysis
presented in section 3.3, of the dependence of the height and
width of this peak on the current and geometry of the strip
(especially the thickness) will be the theme of further research.
This should be helpful to discriminate the different contributions
to the third harmonic response above $T_c$, close to the critical
temperature, i.e.~the thermal oscillation part and
the non-linear conductivity part. \\
\indent Other third harmonic sources (including the non-linear
conductivity) would have to be taken into account to describe more
accurately the whole peak. For example, a temperature shift of the
peak is not yet explained by the proposed theories \cite{MPI},
\cite{MCRI}. Also, slightly above $T_c$, we observe a disagreement
between our fit and the experimental data in the tail of the peak.
Let us discuss in brief some mechanisms for the generation of
harmonics in different temperature regimes and physical
situations, which would have to be taken into account to describe
the whole peak with a better accuracy. \\
\indent At very low temperature, in the vortex pinning regime or
in a vortex-free situation, modulation of the superconducting
order parameter creates dissipation and harmonic generation
\cite{Booth99} \cite{Booth00}; the non-linearity is created by the
current dependence of the superfluid density which can be easily
described within the Ginzburg-Landau model \cite{Borovitskaya}.
Non-linear response of the vortex state \cite{Mallozi} can also
create higher harmonics. However, the complexity of the pinning
mechanisms and the generation of a magnetic field by the applied
current makes it very difficult to develop a detailed theory for
the low-temperature tail of the $\lambda$-shaped maximum. For
example, small superconducting grains in a diamond anvil
\cite{Raphael98}, \cite{Raphael00}, or scanning of the surface of
a thin film by a terminated coaxial cable \cite{Pestov} require
the solution of different electrodynamics problems. In our paper,
we used a narrow strip and the vortices are created near the edges
of the sample by the applied current. The generation of vortices
can be significantly reduced if a Corbino geometry is used; it is
necessary that a superconducting film terminates a coaxial
cable.\\
\indent For temperatures slightly above the critical one, there is
significant fluctuation conductivity. Usual ohmic conductivity of
normal carriers is perfectly linear but fluctuation conductivity
is not and can create an observable part of the harmonic
generation \cite{MPI}.\\
\indent For higher temperatures, temperature oscillations can
create harmonics in the voltage response  \cite{Robbes},
\cite{Cheenne}, \cite{MCRI}, \cite{Zharov}. It is this mechanism
which has been given most attention in our theoretical analysis of
the experimental data for T~$>$~$T_c$ (see Fig. \ref{fig:Fit}).\\
\indent For completeness, we would like to mention that
analogously to temperature oscillations well above $T_c$, a quite
universal mechanism for 3$^{rd}$ harmonic generation at microwave
frequencies exists for the superconducting state below $T_c$. The
density of superfluid particles depends on the applied current and
finally the kinetic inductance of the superconducting sample
depends on the current $I=I_0\cos(\omega t)$
\begin{equation}
\label{Inductance} L(I)~\approx~L_0 + L_1I^2.
\end{equation}
$L_0$ gives the linear contribution to the kinetic inductance and
$L_1$ the non-linear one (see ref. \cite{Booth99},
\cite{Booth00}). Then, the general formula for the voltage
response given by
\begin{equation}
\label{VoltageResponse}
V(t) = L(I)\frac{dI}{dt}
\end{equation}
allows to calculate the third harmonic response \cite{Booth99},
\cite{Booth00}:
\begin{equation}
\label{ThirdHarmBooth} V_{3\omega} = -\frac{L_1\omega
I_0^3}{4}\sin(3\omega t)
\end{equation}
It appears in this formula that the third harmonic response
increases linearly with the frequency. This explains that it is
typically observed in the microwave range devices. In our range of
frequencies (up to 20~kHz), this response is not expected
to be observable.\\
\indent Apart from studying the interface heat resistance, for the
fundamental physics of superconductivity  of layered cuprates it
will be very interesting to investigate the bulk non-ohmic
conductivity related to the kinetics of the normal phase. For
further theoretical
discussion see for example \cite{Hlubina}, \cite{Sandeman} and \cite{HlubinaRice}.\\
\indent In conclusion, the investigation of the $\lambda$-peak of
the third harmonic response is a good qualitative method for the
detection of the superconducting phase transition. Because of the
simplicity of the measurement, it can become a standard method in
the materials science of superconductors and also in the industry
for testing the quality of the thin films and cables. In order to
separate the bulk and interface mechanisms for the third harmonic
generation it is necessary to perform detailed investigations of
superconducting films with different thicknesses. In this way
applied research related to the development of superconducting
bolometers can become an indispensable tool for the investigation
of the fundamental effects in cuprate superconductors
\cite{Gildemeister99} and \cite{Gildemeister00}. Our work
demonstrates that the maximum of the $\lambda$-shaped peak as a
function of temperature can be very easily observed, even without
any special care for signal cleaning.\\
\indent Because the voltage generator produces some third harmonic
signal, if we wish to investigate the third harmonic far from
$T_c$, we have to compensate the parasite 3$\omega$-signal not
only at the level of the data analysis, but also electronically.
However, a very simple filter will lead to a reduction of the
ratio $I_{3\omega}$/$I_0$ ($\approx$ 1/500) by a factor of about
10 only, so that the remaining parasite contribution must still be
taken into account in the data analysis, as described in the
theoretical works (equations (7.10) from \cite{MPI} and (3.8) from \cite{MCRI}).\\
\indent Detailed investigation of the third harmonic response
provides a reliable method for the investigation of superconductor
thermal properties like thermal conductivity, thermal boundary
resistance and heat capacity. Moreover, if the thermal effects are
carefully subtracted, the investigation of the third harmonic
signal will provide the opportunity to extract the relaxation time
of the superconducting order parameter from the isothermal
electric non-linearity. If a Corbino geometry does not create
significant heating of the sample, this can be expected to be the
best experimental setup for the observation of Cooper-pair
depairing non-linearity.\\

{\noindent \bf Acknowledgments}\\

Special thanks goes to Patrick Wagner and Johan~Vanacken
\cite{Wagner} who provided us with YBCO and LSCO thin films. Also
we would like to express our gratitude to Johan~Vanacken for
assistance with the experiments, and to Gunther~Rens for Atomic
Force Microscopy characterisation of the film thickness. T.M.
would like to thank Valya~Mishonova for bringing
ref.~\cite{Zharov} to his attention and for Richard~Hlubina and
Didier~Robbes for their correspondence.

~
\newpage


\end{document}